\documentclass{article}
\usepackage[utf8]{inputenc}

\usepackage{hyperref}
\hypersetup{
    colorlinks = true,
    citecolor = {blue},
    urlcolor = {blue},
}

\usepackage{setspace}
\usepackage{authblk}
\usepackage{changepage} 
\usepackage[breakable]{tcolorbox}
\usepackage{float} 
\usepackage{enumitem}
\title{A study on the Friedmann like Universe with Torsion using Noether Symmetry}
\author{Ramkumar Radhakrishnan}
\affil {Department of Applied Physics, S.V.National Institute of Technology, Surat}
\date{}
\begin{document}
\maketitle
\begin{abstract}
This paper deals with the symmetry analysis of the Einstein Cartan theory which is an extension of the General Relativity and it accounts for the space-time torsion. We begin by applying Noether Theorem to the  Lagrangian of the FRW type cosmology with torsion and choose a point transformation: $(a,\phi,N)\rightarrow(u,v,W)$, such that one of the transformed variable is cyclic for the Lagrangian. Then using the conserved charge, which is obtained by applying Noether theorem, and the constant of motion, we get the solutions and conclude that due to the presence of torsion the FRW type cosmology is in the de Sitter phase.
\end{abstract}
\section{Introduction}
\label{S:1}
The current description of gravity as the geometric property of space-time is given by Einstein's General Theory of Relativity. This theory relates the curvature of space-time to the energy, the momentum of matter and radiation present in that system. In 1922 Friedmann developed a model universe which is known as the Friedmann Universe. This model is governed by a set of equations called the Friedmann equations, which are derived from Einstein's field equations and they describe how the universe contracts or expands, thereby abolishing the idea of a forever static universe. According to him, big bang is followed by expansion, then contraction and an eventual big crunch. Thus he proposed three different models of the universe: Closed universe, Open universe and Flat universe. Also by the observational data of Edwin Hubble and Georges Lemaitre, Einstein accepted  the concept of expanding universe by the combination of the astronomical data with General Relativity. \\\\
Einstein Cartan theory is an extension of Einstein's general theory of relativity which accounts for a space-time torsion. The theory proposed that the torsion is a macroscopic manifestation of the intrinsic spin of matter. This theory was not accepted initially but during the 1960s Kibble and Sciama independently reintroduced the spin of the matter term in the General Theory of Relativity. Technically, torsion or spin or intrinsic angular momentum is defined as the anti symmetric part of the non-Riemannian Affine connection. Thus in the field equations in addition to the metric tensor, there exists a term which corresponds to the torsional field of the matter which in turn corresponds to the total Gravitational force of the universe. The curvature is bent due to space-time and it twists due to the presence of torsion. Considering the dynamic case, the presence of space-time torsion is due to the intrinsic angular momentum of the matter. The field equations with the addition of the torsion field is known as the `Einstein Cartan field equations'.\\\\
Although there is no observational evidence to show us the existence of space-time torsion, still the existence of space-time torsion may resolve many unsolved mysteries in cosmology and astrophysics like singularity, acceleration rate and inflation scenarios of the cosmos.\\\\
The present work deals with modified Friedmann equations in FLRW model in the presence of torsion field and the solutions are evaluated using Noether symmetry of the physical system, and the solutions obtained in turn show that the presence of torsion can modify the standard Friedmann cosmology. Finally, the solutions are analyzed from cosmological view point.
\section{FRW Model with Torsion}
\label{S:2}

 The theory of coupling torsion with gravity is known as Einstein-Cartan theory. In this theory asymmetric affine connection of the space-time is considered , which is compared with the symmetric Christoffel symbols of (pseudo) Riemannian  space in Einstein gravity in the present work. Mathematically, torsion is described as skew-symmetric part of the affine connection as
\begin{equation}
S^a_{bc}=\Gamma^a_{[bc]}.\label{e1}
\end{equation}
Due to covariantly constant (i.e., $\bigtriangledown_c g_{ab}=0$) nature of the metric tensor one can decompose the affine connection as,
\begin{equation}
\Gamma^a_{bc}=\tilde{\Gamma}^a_{bc}+K^a_{bc},\label{e2}
\end{equation}
where $\tilde{\Gamma}^a_{bc}$ (symmetric in $b$ and $c$) is the usual Christoffel symbol\cite{c22} and
\begin{equation}
K^a_{bc}=S^a_{bc}+2S^a_{(bc)}.\label{e3}
\end{equation}
is known as contorsion tensor.
The usual homogeneous and isotropic FLRW\cite{c23} space-time model is considered in the present work as the background space-time with the following line-element:
 \begin{equation}
 ds^2=-dt^2+a^2(t) \Bigg[\frac{dr^2}{1-Kr^2}+r^2 (d\theta^2+\sin^2 \theta~d\phi^2)\Bigg]. \label{e4}
 \end{equation} 
 Here $a(t)$ is the scale factor\cite{c16} and $K=0, \pm 1$ is the curvature index, indicating flat ($K=0$), closed ($K=+1$) and open ($K=-1$) $3D$ hyper surface.\\
 
 It is to be noted that a general form of the torsion field \cite{c18}can not be accommodated in the maximally symmetric FLRW model\cite{c26}. However, keeping the symmetry of the $3D$ space-like hyper surface one can choose the torsion field \cite{c25} as (anti symmetric tensor)
 \begin{equation}
 S_{abc}=2\phi h_{a[b}u_{c]},\label{e5}
 \end{equation}
 where $h_{ab}$ is the metric on the 3-space and $u_c$ is the velocity vector field along the tangent to a congruence of time-like curves.\\
 Now contracting the above torsion tensor\cite{c30}, one gets
  \begin{equation}
  S_{a}=S^b_{ab}=-3\phi u_a,\label{e6}
  \end{equation}
 where $S_a$ is defined as the torsion vector. Clearly torsion vector \cite{c18}is a time-like vector, indicating future directed (if $\phi<0$) or past directed ($\phi>0$).\\
 
 Now, in the presence of torsion,\cite{c30} the matter conservation equation in FLRW model takes the form 
 \begin{equation}\label{e7}
 T_{a;b}^b=-4\phi T_{ab}u^b,
 \end{equation}
 which for perfect fluid reduces to
 \begin{equation}\label{e8}
 \dot{\rho}+3(H+2\phi)(p+\rho)=4\phi\rho.
 \end{equation}
 Here $(\rho, p)$ are the energy density and thermodynamic pressure of the perfect fluid with equation of state $p=w\rho$, where $w$ is a constant. Also, the modified Friedmann equation (with torsion) in the given space-time model has the explicit form\cite{c1}
 \begin{eqnarray}
 3H^2&=&\kappa \rho-3\frac{K}{a^2} -12\phi^2-12H\phi, \label{e9}\\
 2\dot{H}&=& -\kappa (p+\rho) -4\dot{\phi}+8\phi^2+4H\phi. \label{e10}
 \end{eqnarray} 
 \section{Noether symmetry and space-time torsion}
For a given point-like Lagrangian, $L\bigg(q^{\alpha}(x^i),\partial_jq^{\alpha}(x^i)\bigg)$, the usual Euler-Lagrange equations take the form: 
\begin{equation}
\partial
_j\Bigg(\frac{\partial L}{\partial \partial_j q^{\alpha}}\Bigg)-\frac{\partial L}{\partial q^{\alpha}}=0.\label{e11}
\end{equation}
where $q^{\alpha}(x^i)$ are the generalized co-ordinates of the system. For an arbitrary field of function $u^{\alpha}(q^j)$ \cite{c17}the contraction of the above Euler Lagrange  equation is simplified as
\begin{equation}
\bigg(\partial_j \mu^{\alpha}\bigg)\bigg(\frac{\partial L}{\partial \partial_j q^{\alpha}}\bigg)+\mu^{\alpha}\frac{\partial L}{\partial q^{\alpha}}=\partial_j\bigg\{\mu^{\alpha}\frac{\partial L}{\partial \partial_j\partial q^{\alpha}}\bigg\}.\label{e12}
 \end{equation}
Now, corresponding to a vector field in the configuration space\cite{c20}
 \begin{equation}
 \overrightarrow{X}=\bigg(\partial_j \mu^{\alpha}\bigg)\bigg(\frac{\partial }{\partial \partial_j q^{\alpha}}\bigg)+\mu^{\alpha}\frac{\partial }{\partial q^{\alpha}},\label{e13}
 \end{equation}
if one imposes the Noether symmetry to the given physical system by Noether's theorem, then, \cite{c17}
 \begin{equation}
\mathcal {L}_{\overrightarrow{X}}L=\bigg(\partial_j \mu^{\alpha}\bigg)\bigg(\frac{\partial L}{\partial \partial_j q^{\alpha}}\bigg)+\mu^{\alpha}\frac{\partial L}{\partial q^{\alpha}}=0.\label{e14}
\end{equation}
So, from the contraction equation (\ref{e12}) there exists a conserved charge $Q^i$ (with $\partial_i Q^i=0$), namely,\cite{c17}
 \begin{equation}
 Q^i=\Bigg(\mu^\alpha\frac{\partial L}{\partial\partial_i q^{\alpha}}\Bigg),\label{e15}
 \end{equation}
associated with the physical system.\\ Further, the total energy of the system is given by\cite{c17}
\begin{equation}
E=\frac{\partial L}{\partial {\dot{q}}^\alpha}\dot{q}^\alpha-L,\label{e16}
\end{equation}
and it is also a constant of motion.\\\\

In the present gravity theory with torsion in the maximally symmetric FLRW space-time background, the Lagrangian is given by\cite{c19}
\begin{equation}
L(a,\dot{a},\phi,\dot{\phi})=\frac{6a\dot{a}^2}{N}+{6aKN}+{12a^3\dot{\phi}}+{36a^2\dot{a}\phi}+{24{a}^3{\phi}^2N}+{6N\rho_{0}a^{-3w}},\label{e17}
\end{equation}
with $w$ to be constant.\\
Now, the infinitesimal generator of the Noether symmetry, a vector field in the configuration space, has the explicit form \cite{c16}
\begin{equation}
    \overrightarrow{X}=\alpha\frac{\partial}{\partial a}+\beta\frac{\partial}{\partial\phi}+\dot{\alpha}\frac{\partial}{ \partial\dot{a}}+\dot{\beta}\frac{\partial}{\partial\dot{\phi}}+\gamma\frac{\partial}{\partial N},\label{n7.1}
\end{equation}\\
where $\alpha$,~$\beta$,~$\gamma$,~are the functions of augmented space variables i.e., $\alpha=\alpha(a,\phi)$,~$\dot{\alpha}=\frac{\partial\alpha}{\partial a}\dot{a}+\frac{\partial\alpha}{\partial\phi}\dot{\phi}$~and so on.\\By Noether theorem\cite{c16} $\mathcal {L}_{\overrightarrow{X}}L=0$, we have the following partial differential equations for the flat space time
\begin{equation}
\frac{6\alpha}{N}-\frac{6\gamma a}{N^2}+\frac{12a}{N}\frac{\partial\alpha}{\partial a}=0 \label{n8}
\end{equation}
\begin{equation}
12a\frac{\partial\alpha}{\partial\phi}=0
\end{equation}
\begin{equation}
{3a^2\alpha}+3\phi a^2\frac{\partial\alpha}{\partial\phi}+a^3\frac{\partial\beta}{\partial\phi}=0 \label{n8.1}
\end{equation}
\begin{equation}
6\phi a \alpha +3\phi a^2 \frac{\partial\alpha}{\partial a}+3 a^2 \beta +a^3 \frac{\partial\beta}{\partial a}=0
\end{equation}
\begin{equation}
-3\omega\rho_{0}a^{-3w-1}N\alpha+\gamma \rho_{0}a^{-3w}+12\phi^2\alpha a^2N+4\phi^2a^3\gamma+8\phi a^3N\beta=0.\label{n11}
\end{equation}
By solving the above set of partial differential equations we get the value of the unknown functions. And the coefficient of the infinitesimal generator have to satisfy the above equations, which on solving, we get the coefficient of infinitesimal generators as,
\begin{equation}
\alpha=c a,~~\beta=-3c\phi,~~\frac{\gamma}{N}=3c,~~\omega=1 \label{n13.1}
\end{equation}
where c is an arbitrary constant and for simplicity we consider the value of c=1, thus,
\begin{equation}
\alpha= a,~~\beta=-3\phi,~~\frac{\gamma}{N}=3,~~\omega=1 \label{n13.2}
\end{equation}
As Noether symmetry is a particular case of Lie symmetry, the Lie algebra of the Noether symmetry is a sub algebra of the Lie algebra of the Lie symmetry and, one has the conserved charge \cite{c14}
\begin{equation}
Q=\alpha\frac{\partial L}{ \partial\dot{a}}+\beta\frac{\partial L}{\partial\dot{\phi}}\label{n7}
\end{equation}\\
Now due to  the vector field $\overrightarrow{X}$ is transformed into\cite{c20}
\begin{eqnarray}
\overrightarrow{X}_T&=&(i_{\overrightarrow{x}}du)\frac{\partial}{\partial u}+(i_{\overrightarrow{x}}dv)\frac{\partial}{\partial v}+(i_{\overrightarrow{x}}dW)\frac{\partial}{\partial W}
\Bigg(\frac{d}{dt}(i_{\overrightarrow{x}}du)\Bigg)\frac{d}{d\dot u}\nonumber\\
&+&\Bigg(\frac{d}{dt}(i_{\overrightarrow{x}}dv)\Bigg)\frac{d}{d\dot v}+\Bigg(\frac{d}{dt}(i_{\overrightarrow{x}}dW)\Bigg)\frac{d}{d\dot{W}},\label{n15} 
\end{eqnarray}
so that $\overrightarrow{X}_T$ can be considered as the lift of the vector field defined on the augmented space. Further $\overrightarrow{X}$ is a symmetry vector and hence we can restrict the above point transformation such that,
 \begin{equation}
 i_{\overrightarrow{x}}du=1,i_{\overrightarrow{x}}dv=0,i_{\overrightarrow{x}}dW=0,
 \end{equation} where $i_{\overrightarrow{X}}$ is the inner product operator of $\overrightarrow{X}$. 
Therefore $\overrightarrow{X}_T$ takes the simplified form  $\overrightarrow{X}_T=\frac{\partial}{\partial u}$~and~$\frac{\partial L}{\partial u}=0$. This shows that $u$ is a cyclic co-ordinate \cite{c21} and the dynamics of the system can be reduced. Thus we get the relation between the initial variable and the transformed variable as,
 \begin{equation}
e^W=a^3 \phi,~e^v=\phi N,~e^u=a\phi N\label{n15.1}
\end{equation}
The transformed Lagrangian takes the form ,
\begin{eqnarray}
L&=6(\dot{u}-\dot{v})^2e^{W-v}+12\dot{W}e^W+24e^{W+v}+6\rho_{0}e^{v-W},     
\end{eqnarray}
and the conserved charge  in new variables has the form, 
\begin{equation}
Q=12(\dot{u}-\dot{v})^2e^{W-v}.
\end{equation}
Using the conserved charge in the evolution equations for the above transformed Lagrangian, we get the explicit solution of the evolution equation as (in the initial and transformed coordinates)\\\\
{\bf For}~~$\omega=1, K=0,c=1$\\
\begin{equation}
u=e^W{\sqrt{\rho_{0}e^{-2v}-4}}t+c_{1}
\end{equation}
\begin{equation}
v=v_0  
\end{equation}
\begin{equation}
W=W_0 
\end{equation}
\begin{equation}
a= e^{(e^{W_0}{\sqrt{\rho_{0}e^{-2v_0}-4}}t+c_{1}-W_0)}   
\end{equation}
\begin{equation}
\phi=e^{(v_0+3W_0-3e^{W_0}{\sqrt{\rho_{0}exp(-2v_0)-v_0}}t+c_{1}} \\\\
\end{equation}

where $c_{1},W_0,v_0,\rho_{0}$ are constants.
 \section{Brief Discussion}
 The exponentially increasing value of the above scalar field `$a$', implying the expansion of the Universe (figure 1), describes this model with torsion\cite{c30} to be in de Sitter space in a flat space-time. The presence of torsion can ultimately change evolution of the classical homogeneous and isotropic Friedmann Universe, which means, torsion can lead to an exponential expansion. Thus for this instance, due to the presence of torsion the Friedmann universes evolve like the de Sitter model. Also the results obtained due to the application of the Noether Theorem show that the model is in an inflationary phase in the early Universe whose value of the decelerating parameter is constant which is true for a model represented in de Sitter space in a flat space. Generally the introduction of space-time torsion is incompatible with the high symmetry of the FRW cosmologies and hence the solution obtained is in the de Sitter space.\\\\
  A comparative study can be done by using the value of the Hubble parameter of the Universe without torsion. The universe without the torsion would have either an increasing or decreasing variable value. But the presence of torsion makes the Hubble parameter to have a constant value.\\\\\\
  If the value of `W' is negative it would result in the de sitter deflation whereas a positive value of `W' would result in the de sitter inflation. But at the end the Universe would only exist in the de sitter space due to the presence of torsion.The existence of this model in the de sitter space is analogous to the torsion free counterparts, the interesting nature being, the torsion field takes the very
 restricted form which is imposed by the high symmetry of the Friedmann-like Universe.\\\\
\begin{figure}
 	\centering
 	\includegraphics[width=0.5\textwidth]{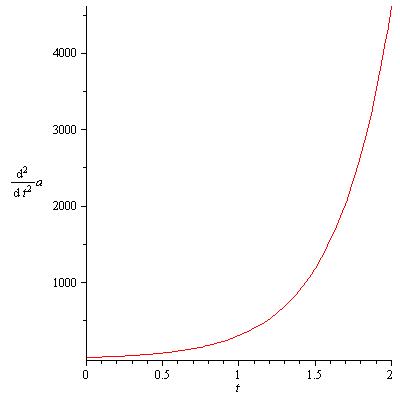}\\
 	\caption{Representation of the acceleration parameter with respect to the cosmic time }
 	\label{fig1}
 \end{figure}
Therefore the present work is an example of the evolution equation being highly coupled differential equations which is very tedious to solve by usual methodology. But the solution can be obtained by applying the Noether symmetry to it and the unknown functions are obtained by the application of symmetry conditions. And also, the results obtained by the application of Noether theorem go hand in hand with the results obtained with the other methods used for finding its nature.
\newpage
\section*{Acknowledgment}
\label{S:5}
I would like to express my special thanks of gratitude to \textbf{Dr.Subenoy Chakraborty}, Professor, Department of Mathematics, Jadavpur University, Kolkata who gave me this opportunity to work on this problem during the time of my visit to Jadavpur University in the Summer 2019. His clarity of thoughts have been present at every moment of this work. I owe a great intellectual debt to him. I thank him for encouraging and helping me to solve the problem.  I also thank \textbf{Dr.Sourav Dutta}, Postdoctoral fellow,  Department of Mathematics, Jadavpur University, Kolkata who worked with me till the end to solve this problem. I also thank my parents without whom this would have not been possible.

\end{document}